\pdfoutput=1 
\documentclass{aa}
\usepackage{graphicx}
\usepackage{txfonts}
\usepackage{color}
\usepackage{tabularx} 
\usepackage{adjustbox}
\usepackage{float}
\usepackage{mydblfloatfix} 
\newcommand{\be}{\begin{equation}}
\newcommand{\ee}{\end{equation}}
\newcommand{\bea}{\begin{eqnarray}}
\newcommand{\eea}{\end{eqnarray}}

\begin{document}

\title{When do solar erupting hot magnetic flux ropes form? 
}
\author{A. Nindos \inst{1}
\and S. Patsourakos \inst{1}
\and A. Vourlidas \inst{2}
\and X. Cheng \inst{3,4}
 \and J. Zhang \inst{5}}
\institute{Physics Department, University of Ioannina, Ioannina GR-45110,
Greece\\
\email{anindos@uoi.gr}
\and
The Johns Hopkins University Applied Physics Laboratory, Laurel, MD 20723, USA
\and
School of Astronomy and Space Science, Nanjing University, Nanjing 210093,
China
\and
Max Planck Institute for Solar System Research, G\"{o}ttingen D-37077, Germany
\and
Department of Physics and Astronomy, George Mason University, Fairfax, VA 
22030, USA}

\date{Received: Accepted:}

 
  \abstract
{}
{We investigate the formation times of eruptive magnetic flux ropes relative to
the onset of solar eruptions, which is important for constraining models of 
coronal mass ejection (CME) initiation.}
{We inspected uninterrupted sequences of 131 \AA\ images that spanned more than
eight hours and were obtained by the Atmospheric Imaging Assembly  (AIA) 
on board the Solar Dynamics Observatory (SDO) to identify the formation times 
of hot flux ropes that erupted in CMEs from locations close to the limb. The 
appearance of the flux ropes as well as their evolution toward eruptions were 
determined using morphological criteria.}
{Two-thirds (20/30) of the flux ropes were formed well before  the onset of the eruption (from 51 minutes
to more than eight hours), and their formation 
was associated with the occurrence of a confined flare. We also found four 
events with preexisting hot flux ropes whose formations occurred a matter of minutes
(from three to 39) prior to the eruptions without any association with distinct 
confined flare activity. Six flux ropes were formed once the eruptions were
underway. However, in three of them,
prominence material could be seen in 131 \AA\ images, which may indicate  the
presence of preexisting flux ropes that were not hot. The formation patterns of the last three groups of hot flux ropes
did not show significant differences. For the whole population of events,
the mean and median values of the time difference between the onset of the 
eruptive flare and the appearance of the hot flux rope were 151 and 98 minutes,
respectively.}
{Our results provide, on average, 
indirect support for CME
models that involve preexisting flux ropes; on the other hand, for a third of the
events, models in which the ejected flux rope is formed during the eruption
appear more appropriate.}


   \maketitle
%

\section{Introduction}

Coronal mass ejections (CMEs) are the most spectacular form of dynamic
phenomena in the solar atmosphere. They are defined as large-scale
expulsions (on the order of $\sim10^{14}$ to $\sim10^{16}$ g) of coronal
magnetized plasma into the heliosphere at speeds  ranging from about
100 km s$^{-1}$ to more than 2000 km s$^{-1}$ (e.g., see Yashiro
et al. 2004; Chen et al. 2011, and references therein).

The preeruptive configuration of CMEs has been extensively
debated. There is a vast literature on the subject, for example  reviews by Forbes (2000), Klichmuk
(2001), Chen (2011), Aulanier (2014),  Schmieder et al. (2015), Cheng
et al. (2017), Green et al. (2018), Georgoulis et al. (2019), and
references therein.
Briefly, there  are two groups of CME models depending on
the configuration of the coronal  magnetic field  before eruption. One
group of models rely on the existence of a magnetic flux rope (i.e., a
twisted flux tube whose magnetic field lines wind about an axial
magnetic field line in the interior of the tube) prior to the eruption
(e.g., Forbes \& Isenberg 1991; Isenberg et al. 1993; Forbes \& Priest
1995; Gibson \& Low 1998; Roussev et al. 2003;  Amari et al. 2004,
2005; T\"{o}r\"{o}k \& Kliem 2005; Kliem \& T\"{o}r\"{o}k 2006; Fan
\& Gibson 2007; Archontis \& T\"{o}r\"{o}k 2008;  Archontis \& Hood
2012). The most convincing observational evidence in favor of models 
with preexisting flux ropes is the identification of a flux-rope-like
structure that is present well before the eruption (references are provided
later in this section, in conjunction with the discussion of proxies to
active region flux ropes).

In the models of the second group, the preeruptive
configuration consists of sheared magnetic arcades, that is to say a set of
arc-like magnetic field lines whose orientation deviates from the
local normal to the polarity inversion line. 
It has been suggested
that magnetic reconnection between two sheared sets of loops within
the core field of a bipolar active region may create (or grow) a flux rope 
that immediately erupts outward (``tether-cutting'' scenario, 
e.g.,
Miki\'{c} \& Linker 1994; Amari et al. 1996; Roussev et al. 2004; Jacobs et al.
2006). Several eruptive events have been interpreted in terms of this model
(e.g., Moore et al. 2001; Liu et al. 2012; Chen et al. 2014b; 2016; Xue et al.
2017); usually the observations include flare brightenings and weak precursor
activity that takes place prior to the eruption at the core of the active region. 
Alternatively, the core of a sheared arcade may expand (for example,
due to shearing motions) and then erupt after it breaks through the
overlying magnetic field via reconnection at a null point (breakout model, 
see Antiochos et al. 1999; MacNeice et al. 2004; Lynch et al. 2008; Karpen et
al. 2012; Lynch \& Edmondson 2013). Observations favoring the breakout model
include eruptions occurring in multipolar magnetic configurations (Aulanier et 
al. 2000; Ugarte-Urra et al. 2007) as well as small-scale brightenings 
that appear away from the active region core (e.g., see Sterling \& Moore 2004 
and references therein). 

Although there is no broad consensus on the state of the preeruptive
magnetic field, practically all models predict that the CME will contain
a flux rope after eruption; in other words, while the flux rope is an
integral part of the preeruptive configuration in the first group of
models,  it forms during the eruption in the second group of
models. Therefore, it is important to know when the flux rope that 
subsequently erupts is formed.

Since the ejected structure is a flux rope, white light coronagraph 
observations of CMEs or in situ solar wind measurements cannot provide a direct
way to determine the type of preeruptive configuration. Searching for 
evidence for the existence of preeruptive flux ropes in the low corona
is complicated by the lack of routine observations of the coronal
magnetic field. Therefore, we usually resort to proxies. In active regions, the 
following four proxies have been used: 
(1) thread orientation in filaments (e.g., L\'{o}pez-Ariste et
al. 2009; Jiang et al. 2014; Yardley et al. 2016; Chandra et al. 2017;
Xue et al. 2016);
(2) soft X-ray sigmoids when their middle section crosses the polarity
inversion line in the inverse direction (e.g., Green \& Kliem 2009;
Green et al. 2011) and when their ends turn around to point toward the center of the S; 
(3) nonlinear force-free field (NLFFF) extrapolations of the
photospheric vector magnetograms (e.g., Yan et al. 2001; Canou et
al. 2009; Guo et al. 2013; Chintzoglou et  al. 2015; Amari et al. 2018; 
James et al. 2018; Duan et al. 2019); 
and (4) hot flux ropes. Regarding this last proxy, high-cadence, high spatial resolution observations
obtained with the Atmospheric Imaging Assembly (AIA; Lemen et
al. 2012) aboard the Solar Dynamics Observatory (SDO; Pesnell et
al. 2012) have revealed the existence of coherent ``hot
channels'' or ``hot blobs'' (Reeves \& Golub 2011; Cheng et al. 2011,
2012, 2013, 2014a, 2014b, 2014c; Zhang et al. 2012; Li \& Zhang 2013;
Tripathi et al. 2013; Patsourakos et al. 2013; Chen et al.  2014a,
2020; Joshi et al. 2014; Song et al. 2014; Vemareddy \&  Zhang 2014;
Nindos et al. 2015; Chintzoglou et al. 2015; James et al. 2017, 2018;
Zhou et al. 2017; Veronig et al. 2018; Wang et al. 2019). These
structures appear in the 131 \AA\ passband, which probes temperatures of
$\sim10$ MK during flares, but they do not appear in cooler passbands
that probe quiescent coronal temperatures ($\lesssim 2$ MK). 

These hot channels and blobs 
have been identified as hot flux ropes
based on morphological criteria (see Sect. 2), which interpret
observations on the grounds of theoretical expectations. Obviously,
this interpretation of observations may have to be modified in the
future if needed. Therefore, it would probably be more appropriate to
use the term ``candidate hot flux ropes'' but for the sake of brevity we
will, instead, adopt the term ``hot flux ropes'' (HFRs) throughout the 
paper.

The appearance of HFRs is quite common in large events; using an
extensive  dataset of 141 M-class and X-class flares that occurred
relatively close to the limb, Nindos et al. (2015;  hereafter referred
to as Paper I) found HFRs in 45 of them (32\%). Furthermore, they
detected HFRs in 34 of their 70 eruptive events (49\%). 

There are several case studies on the formation times of HFRs.  These
studies take advantage of the high-cadence uninterrupted sequence of
full-disk observations provided by the AIA and are also favored by the
fact that the candidate flux ropes usually remain coherent throughout most
of their  evolution. A few events with HFRs forming during
the eruption have been reported (e.g., Cheng et al. 2011; Song et
al. 2014; Veronig et al. 2018). Most publications report preexisting
HFRs, but the times of the reported appearances are biased by the
duration of the observational window. Those that  use rather short
observational windows around the  eruption report the formation of
HFRs a matter of minutes (two to 30) before the onset of the eruption (Cheng
et al.  2013; Li \& Zhang 2013; Tripathi et al. 2013;  
Cheng et al. 2014a,c, Vemareddy \& Zhang 2014; Cheng et al. 2015). 

Patsourakos et al. (2013) were the first to report the formation of an
HFR well  before  its eruption (about seven hours), during a confined
flare. If they had  used a shorter time interval they would have
missed the confined event and  their conclusion about the time of the
HFR formation would have been  different. In subsequent publications,
similar conclusions about the formation of HFRs long before the eruption (from one
to more than 11 hours) were reported by a
number of authors (Chen et al. 2014a; Cheng et al. 2014b; Chintzoglou
et al.  2015; James et al. 2017, 2018; Kumar et al. 2017; Zhou et
al. 2017; Liu et al.  2018; Wang et al. 2019).

The above publications were case studies that did not address the
statistics of the formation times of HFRs in eruptive events. With the
present study, we aim to remedy this issue by determining the formation
times of HFRs in an extensive dataset of large eruptive events using
observations that cover several hours before the onset of the
eruptions. The paper is organized as follows.  In Sect. 2, we present
the observations and data analysis. In Sect. 3, we introduce our
classification of events in terms of the formation times of the
HFRs. In Sect. 4, we discuss each category of events. Finally, in Sect.
5, we discuss our results and conclude. 

\section{Observations and data analysis}

Our starting point was the 34 eruptive M-class and X-class flare
events from Paper I that involved an HFR configuration. We analyzed
each event using extreme ultra-violet (EUV) images 
of the low corona from the AIA. The pixel
size, cadence, and field of view of the AIA images are 0.6\arcsec, 12
s, and 1.3$R_{\odot}$, respectively. We used images obtained  in
narrowband channels centered at 94, 131, 171, 193, 211, 304, and 335
\AA,\ which have peak responses at about 6.3, 10, 0.6, 1.6, 2.0, 0.05,
and 2.5 $\times 10^6$ K, respectively (Lemen et al. 2012). The
emission in the 94 and 131  \AA\ passbands probes multi-million Kelvin
plasma only during flares; during quiescent periods, their signal is
dominated by emissions from cooler plasmas with temperatures below
10$^6$ K (O'Dwyer et al. 2010). In each passband, our data covered an
interval of more than eight hours, ending a few minutes after the
onset time of the eruption.  The study of time intervals much longer than those in Paper I (the maximum duration of the movies we used in Paper I
was one hour) was deemed necessary for the  identification of HFRs
that formed long before the eruptions. 

To reduce  data size, the cadence of the AIA data was degraded to two
minutes. For  selected events, we compared movies made with two-minute
cadences with full-cadence movies and found that the two-minute cadence
was adequate to capture the various dynamics. All images  of each
event were rotated to a common reference time to correct for solar
rotation.

In Paper I, the identification of the flux ropes was done using 131
\AA\ data obtained around the time of the eruptive flares and employing a 
set of morphological criteria.  Briefly, these included the identification of:
(1) ring-like structures or round  blobs interpreted as flux ropes seen
edge-on. These structures usually appeared on top of $\Lambda$-shaped
features, cusps, and current-sheet-like thin elongated features. (2) Tangled threads winding  around a central, axial direction or
either twisted or writhed
structures interpreted as flux ropes  seen face-on. 
Close to disk center, sigmoidal structures were also interpreted as flux
ropes seen face-on.
(3) Configurations with coexisting morphologies of both cases (1) and (2) 
interpreted as flux ropes seen from intermediate viewpoints. 

The same criteria were also used for the identification of 
and for monitoring the evolution of the flux ropes in our
eight-hour-long 131 \AA\ movies. As in Paper I, we ensured that all
candidate flux ropes were hot by their absence in the 171 and
304 \AA\ movies.  

\begin{table*}[h]
\begin{center}
\caption{Timing of HFR eruptive events}
\begin{tabular}{rrrccc}
\#\tablefootmark{a} & Flare start and peak      & HFR appearance    &  $\Delta T$\tablefootmark{b}  & Confined flare associated  & Classification\tablefootmark{c} \\
  &                            &                   & (minutes) &  with HFR appearance  & \\
\hline
5 &   2011 Feb 24 07:23 07:35  & 2011 Feb 24 07:26 & -3         & No  & FLYP \\
8 &   2011 Mar 07 19:43 20:12  & Uncertain         & –         & – & UNC \\
11 &  2011 Mar 08 03:37 03:58  & 2011Mar 08 01:58  & 99        & Yes (1.2, C2.2) & PREC\\
13 &  2011 Mar 08 19:35 20:16  & 2011Mar 08 10:39  & 536       & Yes (38.3, M5.3) & PREC \\
23 &  2011 Sep 10 07:18 07:40  & 2011 Sep 10 05:28 & 110       &  Yes* (1.09) & PREC \\
25 &  2011 Sep 22 10:29 11:01  & 2011 Sep 22 08:51 & 98        & Yes (5.6, C8.9) & PREC \\
35 &  2011 Oct 22 10:00 11:10  & 2011 Oct 22 09:41 & 19        & No & PRE \\
49 &  2012 Jan 27 17:37 18:36  & 2012 Jan 27 16:44 & 53        & Yes (1.3, B8.8) & PREC \\
52 &  2012 Mar 04 10:29 10:52  & 2012 Mar 04 03:31 & 418       & Yes (4.0, C2.8) & PREC \\ 
54 &  2012 Mar 13 17:12 17:30  & Uncertain         & –         & – & UNC \\ 
61 &  2012 May 17 01:25 01:47  & 2012 May 17 00:46 & 39        & No & PRE \\
65 &  2012 Jul 08 09:44 09:53  & Uncertain         & –         & – & UNC \\ 
67 &  2012 Jul 08 16:23 16:32  & 2012 Jul 08 10:27 & 356       & Yes (8.0, C6.9) & PREC \\
69 &  2012 Jul 19 04:17 05:58  & 2012 Jul 18 22:05 & 372       & Yes (8.3, C4.5) & PREC \\
70 &  2012 Jul 27 17:17 17:26  & 2012 Jul 27 17:22 & -5        & No & FLY \\
81 &  2012 Nov 08 02:08 02:23  & 2012 Nov 07 22:51 & 197       & Yes* (1.2) & PREC \\
87 &  2013 Mar 21 21:42 22:04  & 2013 Mar 21 20:06 & 96        & Yes* (1.7) & PREC \\
89 &  2013 May 03 17:24 17:32  & 2013 May 03 17:26 & -2         & No & FLY \\
94 &  2013 May 13 15:48 16:05  & 2013 May 13 12:10 & 218       & Yes (11.7, M1.3) & PREC \\
95 &  2014 May 14 01:00 01:11  & 2013 May 13 21:45 & 195       & Yes (6.8, C8.3) & PREC \\
96 &  2013 May 15 01:25 01:48  & 2013 May 14 18:31 & 414       & Yes (2.5, C2.0) & PREC \\
97 &  2013 May 20 05:16 05:25  & 2013 May 20 03:49 & 86        & Yes (1.6, B7.9) & PREC \\
98 &  2013 May 22 12:35 13:32  & 2013 May 22 12:37 & -2        & No & FLY \\
99 & 2013 Jun 07 22:32 22:49   & 2013 Jun 07 22:35 & -3        & No & FLYP \\
100 & 2013 Jun 21 02:47 03:14  & 2013 Jun 20 23:51 & 176       & Yes (0.6, B7.1) & PREC \\
103 & 2013 Oct 25 02:48 03:02  & 2013 Oct 25 00:49 & 119       & Yes (1.1, C2.1) & PREC \\
107 & 2013 Oct 26 19:22 19:27  & 2013 Oct 26 19:18 & 4         & No & PRE \\
108 & 2013 Oct 27 12:36 12:48  & Uncertain         & -         & - & UNC \\
109 & 2013 Oct 28 01:41 02:03  & 2013 Oct 28 01:38 & 3         & No & PRE \\
115 & 2013 Nov 21 10:52 11:11  & 2013 Nov 21 10:01 & 51        & Yes (1.1, C1.2) & PREC \\
117 & 2014 Jan 04 22:09 22:52  & 2014 Jan 04 21:10 & 62        & Yes* (0.5) & PREC \\
118 & 2014 Jan 08 03:39 03:47  & 2014 Jan 07 21:43 & 356       & Yes* (0.9) & PREC \\ 
130 & 2014 Feb 09 15:40 16:17  & 2014 Feb 09 15:38 & 0         & No & FLYP \\ 
133 & 2014 Feb 24 11:03 11:17  & 2014 Feb 24 03:32 & 451       & Yes (2.4, C2.4) & PREC \\
\hline 
\end{tabular}
\tablefoot{In the fifth column, asterisks denote events for which the confined 
flare has been detected only in the AIA data and not in the GOES data. Values in 
parentheses give the peak of the 131 \AA\ confined flare emission in millions of 
DN s$^{-1}$ and, when appropriate, its GOES classification.  
\tablefoottext{a}{Event number as registered in Paper I.}
\tablefoottext{b}{$\Delta t$ denotes the time difference between the onset of
the eruptive flare and the appearance of the HFR.}
\tablefoottext{c}{The acronyms FLY, FLYP, PRE, PREC, and UNC are 
explained in Sect. 3.}
}
\end{center}
\end{table*}

In Paper I, the existence of HFRs in 131 \AA\ was further confirmed by
the identification of flux rope structures in Hinode X-ray Telescope
(XRT) data.  The response of all XRT filters is broad (see Narukage et
al. 2011), which allows them to probe material at a broad temperature
range (from above 2 MK to more than 10
MK for the thin filters). Therefore, as we noted in Paper I, looking at thin-filter XRT
images can be considered as roughly equivalent to looking at images
from the narrowband 211, 335, and 94 \AA\ AIA passbands. However, only ten
of the 34  eruptive AIA events were also observed by the XRT, and flux
rope morphologies  were detected in seven of them. Furthermore, it was
difficult to find uninterrupted eight-hour-long sequences of XRT data for
any of them. For these reasons, instead of checking XRT data for the
long-term evolution of flux ropes, we checked
eight-hour-long sequences of  images from 94, 193, 211, and 335
\AA\ AIA passbands for this paper. All HFRs identified in the 131 \AA\ data appear in 94
\AA\ as well, but only 12 of them appear in some of the cooler  passbands centered at
211 and 335 \AA\ (and these appear fuzzier and for
limited time intervals). 
In agreement with the results of Paper I, none of
them appeared in the eight-hour-long movies of the coolest passbands
centered at 171 and 304 \AA.

Inspired by the work of Patsourakos et al. (2013), who reported the
formation of an HFR during a confined flare several hours before the
eruption of the flux rope, we searched for confined flares associated
with the appearance of the HFRs that later erupted. As will be seen in
Sect. 3, that correlation (or the absence thereof) was an important 
ingredient of the classification scheme that we adopted. To this end, for
each event we constructed light  curves in all AIA channels that we
used by integrating the emission (normalized to unity exposure time) within 
a box containing the HFR after its appearance. 

\begin{figure*}[h!]   
\centering
\includegraphics[width=0.70\textwidth]{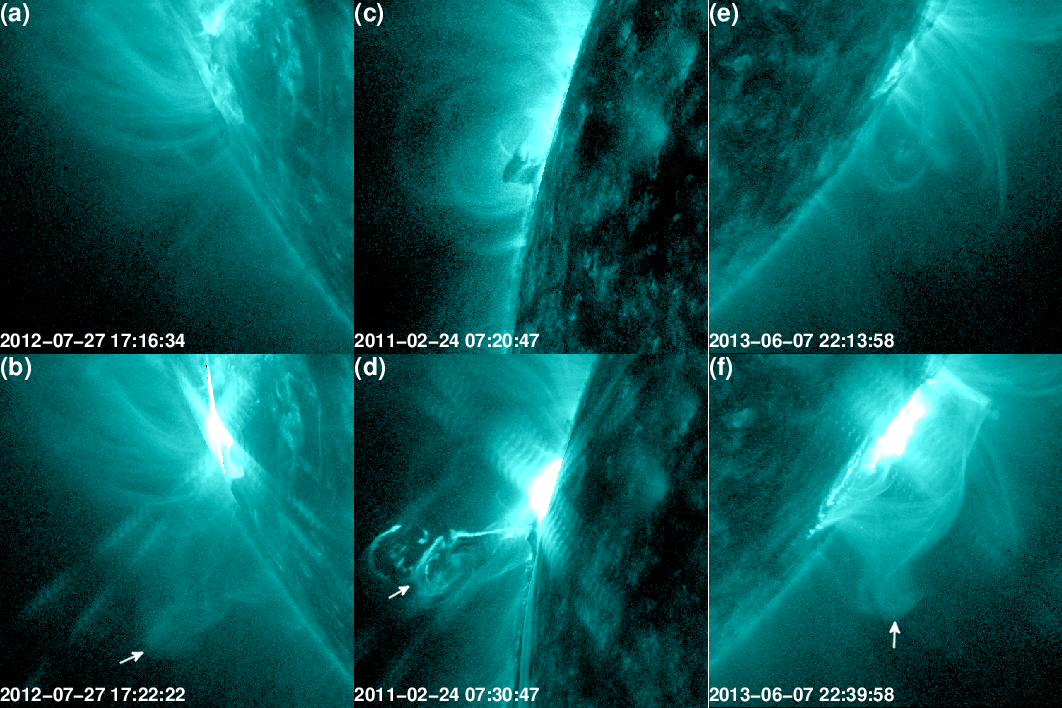}
\caption{Events, at 131 \AA, with HFRs formed during the eruption 
with (middle and right-hand columns; FLYP) and without (left column; FLY) preexisting prominence material. Top row: snapshots of the pre-event 
configuration. Bottom row: snapshots after the appearance of the HFRs. The 
images of the left-hand, middle, and right-hand columns correspond to events 70, 
5, and 99 in Table 1, respectively. The arrows mark the HFRs. The field of 
view in each image is $300 \times 300$ arcsec$^2$.} 
\end{figure*}

\begin{figure*}[h!]
\centering
\includegraphics[width=0.80\textwidth]{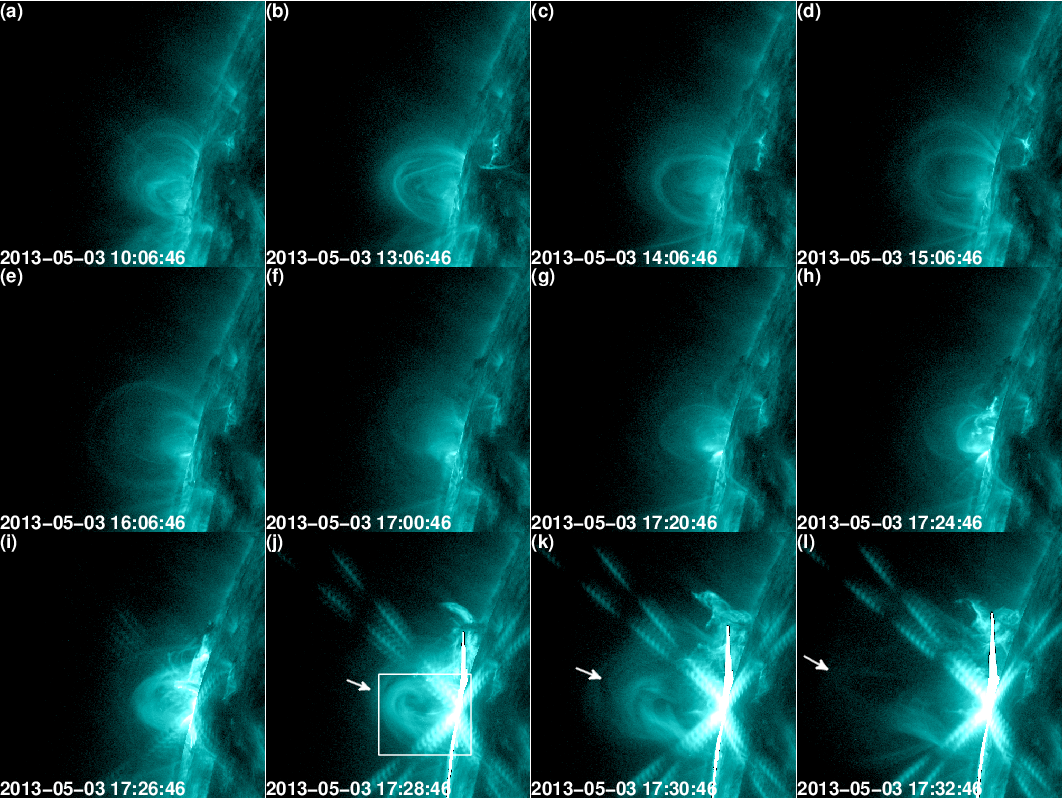}
\caption{Example of the formation, at 131 \AA, of an HFR during the
eruption (FLY; event 89 in Table 1). The arrows mark the HFR. The box in panel
(j) marks the area used  for the calculations presented in
Fig. 3. The field of view is $300 \times 300$ arcsec$^2$. See also the 
associated movie.}
\end{figure*}

\begin{figure}
\centering
\includegraphics[width=0.50\textwidth]{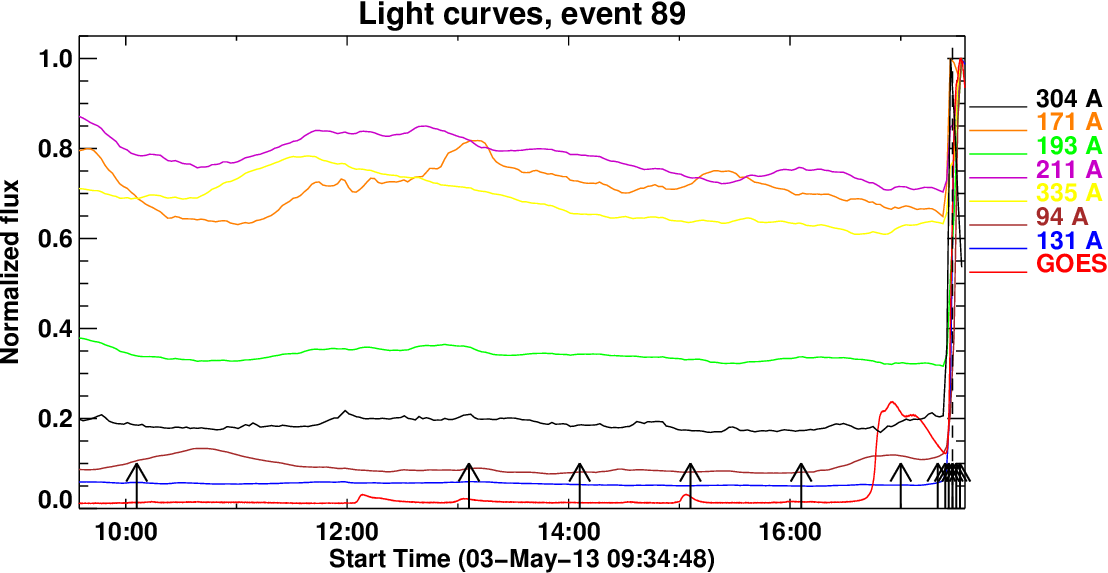}
\caption{Light curves in 94, 131, 171, 193, 211, 304, and 335 \AA\ AIA 
passbands from the region marked with the box shown in Fig. 2(j) as well as
in the GOES 1-8 \AA\ channel. All curves are normalized to their maximum values.
The vertical dashed line marks the first appearance of the HFR.
The arrows just above the horizontal axis mark the times of the images 
displayed in Fig. 2.}
\end{figure}

\begin{figure}[h!] 
\centering
\includegraphics[width=0.45\textwidth]{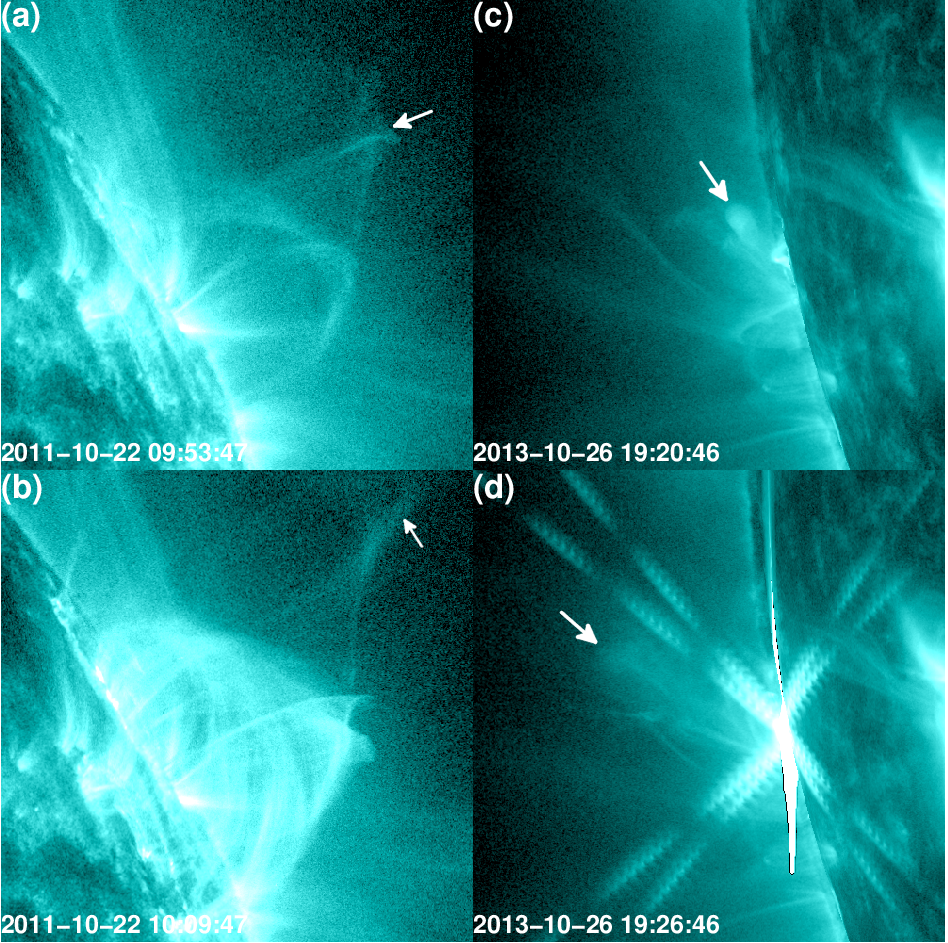}
\caption{Events, at 131 \AA, with preexisting HFRs whose formations were not 
associated with confined flares (PRE). Top row: snapshots of HFRs observed a 
few minutes after their appearance. Bottom row: the flux ropes of the top row 
observed at a later stage. The images of the left- and right-hand columns correspond 
to events 35 and 107 in Table 1, respectively. The arrows mark the HFRs. The 
field of view in each image is $300 \times 300$ arcsec$^2$.}
\end{figure}

\begin{figure*}[btp!] 
\centering
\includegraphics[width=0.80\textwidth]{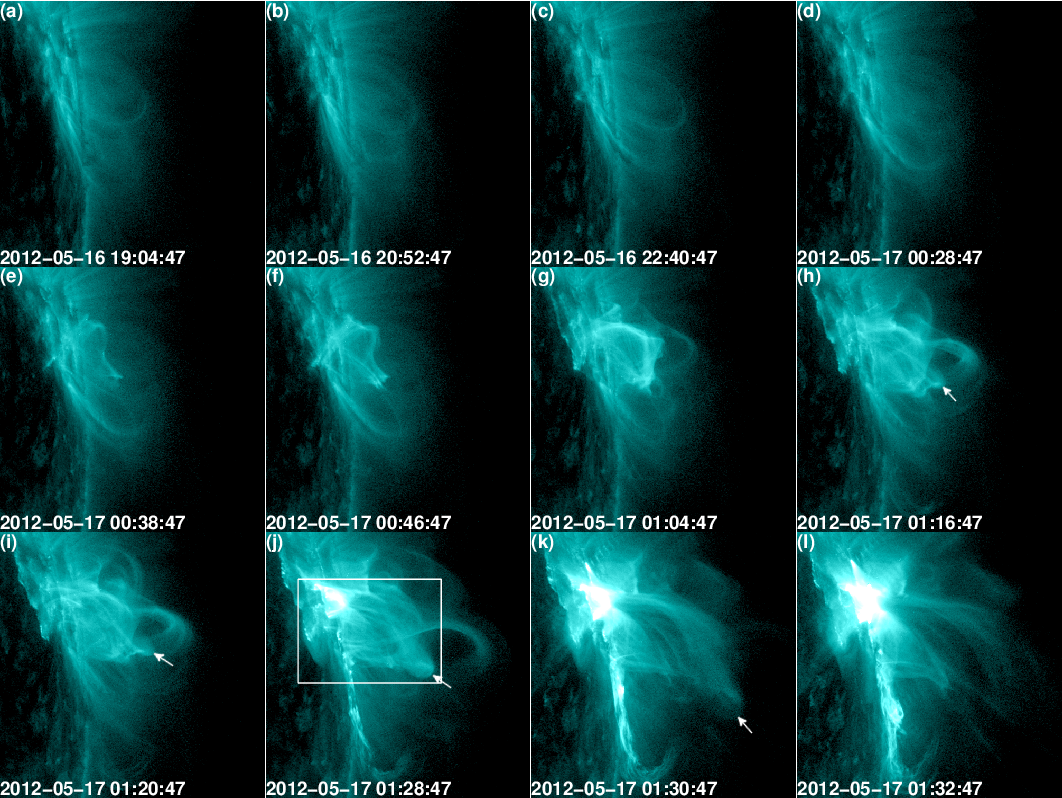}
\caption{Example of the evolution, at 131 \AA, of a preexisting HFR whose 
formation was not associated with a confined flare (PRE; event 61 in Table 1). 
The arrows mark the HFR. The box in panel (j) marks the area used for the 
calculations presented in Fig. 6. The field of view is $300 \times 300$ 
arcsec$^2$. See also the associated movie.}
\end{figure*}

\begin{figure}[h!] 
\centering
\includegraphics[width=0.50\textwidth]{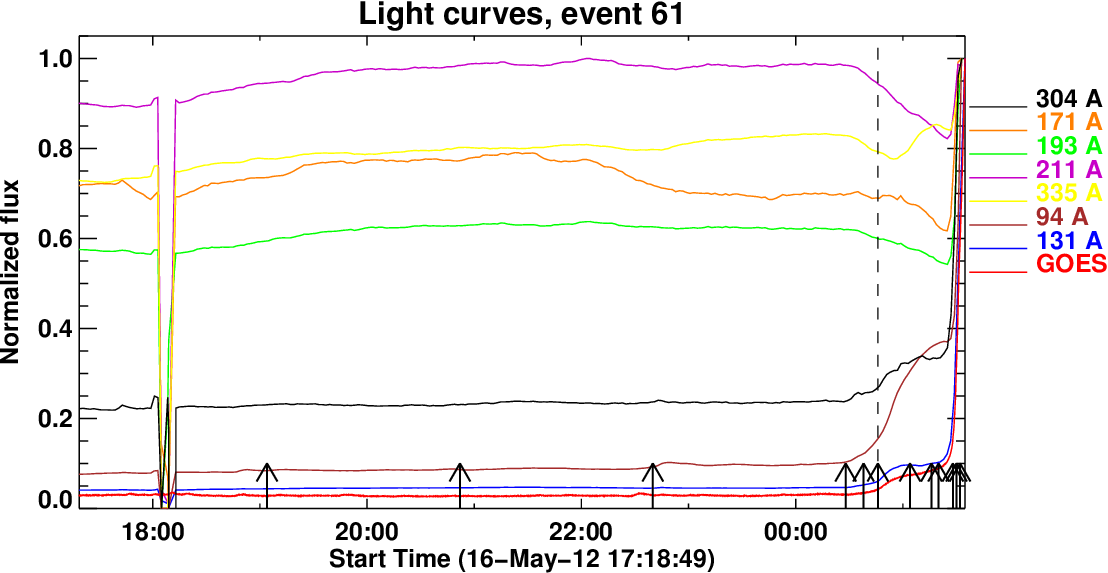}
\caption{Same as Fig. 3 but for the event presented in Fig. 5.}
\end{figure}

\begin{figure*}[h!] 
\centering
\includegraphics[width=0.70\textwidth]{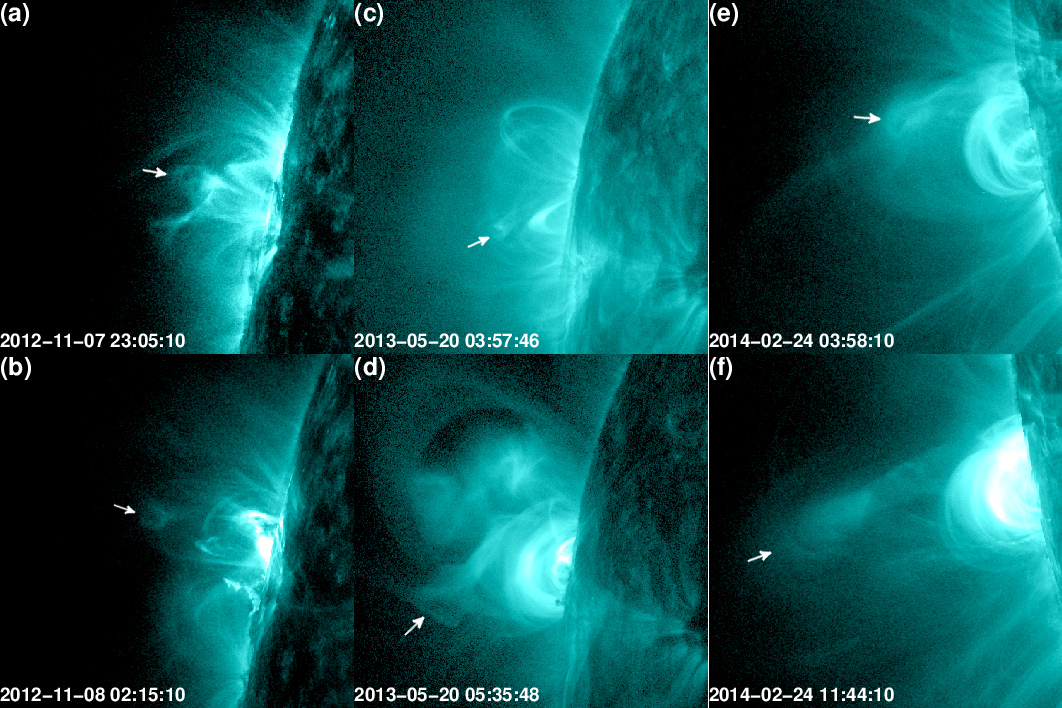}
\caption{Same as Fig. 4 but for preexisting HFRs whose formations were 
associated with confined flares (PREC). The images in the left-hand, middle, and 
right-hand columns correspond to events 81, 97, and 133 in Table 1, respectively. 
The field of view in each image is $300 \times 300$ arcsec$^2$ except in 
those of the middle column where it is $360 \times 360$ arcsec$^2$.}
\end{figure*}

\begin{figure*}
\centering
\includegraphics[width=0.80\textwidth]{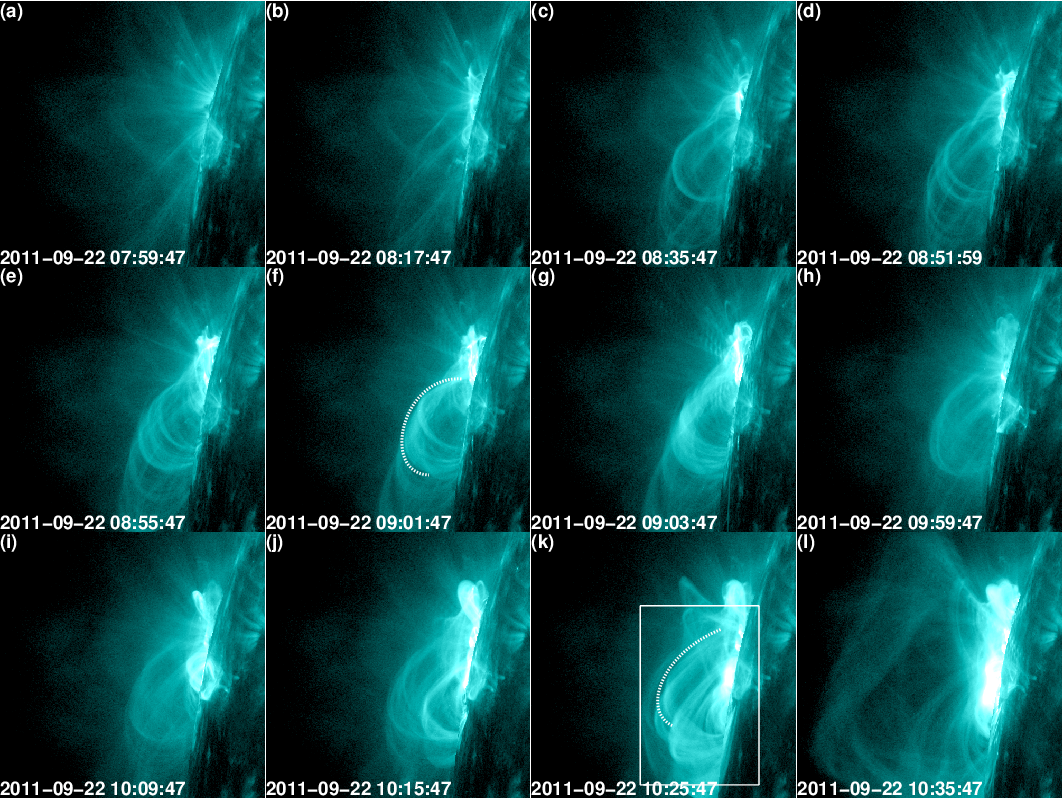}
\caption{Example of the evolution, at 131 \AA, of a preexisting HFR whose 
formation was associated with a confined flare (PREC; event 25 in Table 1). 
The dotted curves delineate the outer edge of the HFR. The box in panel (k) 
marks the area used for the calculations presented in Fig. 9. The field of 
view is $300 \times 300$ arcsec$^2$. See also the associated movie.}
\end{figure*}

\begin{figure}[h!]
\centering
\includegraphics[width=0.50\textwidth]{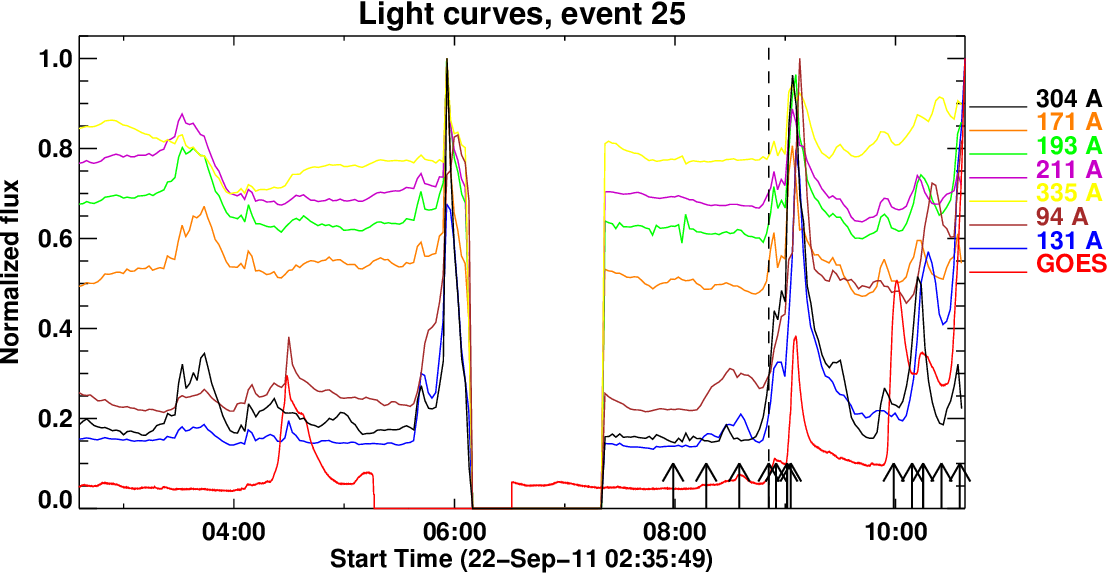}
\caption{Same as Fig. 3 but for the event presented in Fig. 8.}
\end{figure}

\begin{figure*}
\centering
\includegraphics[width=0.80\textwidth]{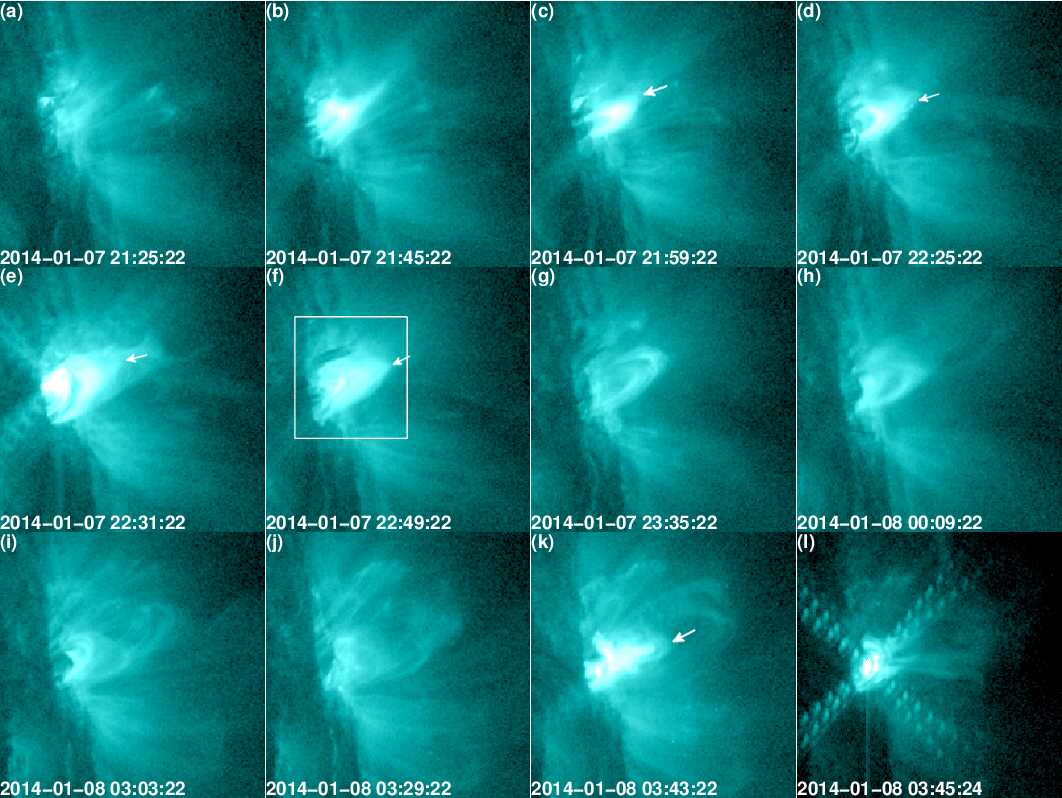}
\caption{Same as Fig. 8 but for event 118 in Table 1. The arrows mark the 
HFR. The field  of view is $180 \times 180$ arcsec$^2$. See also the associated
movie.}
\end{figure*}

\begin{figure}
\centering
\includegraphics[width=0.50\textwidth]{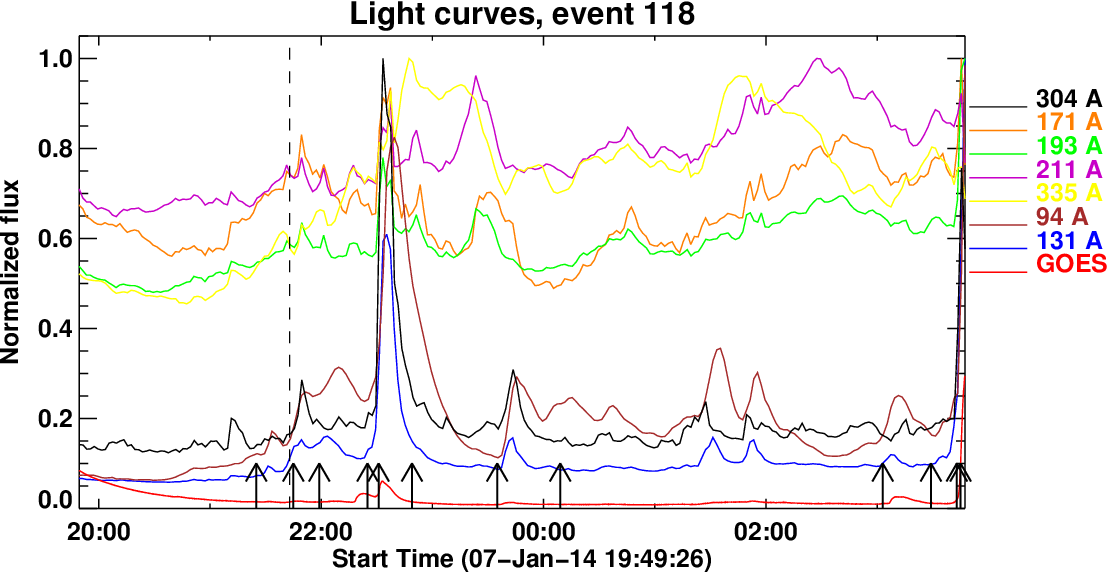}
\caption{Same as Fig. 3 but for the event presented in Fig. 10.}
\end{figure}

\begin{figure}
\centering
\includegraphics[width=0.50\textwidth]{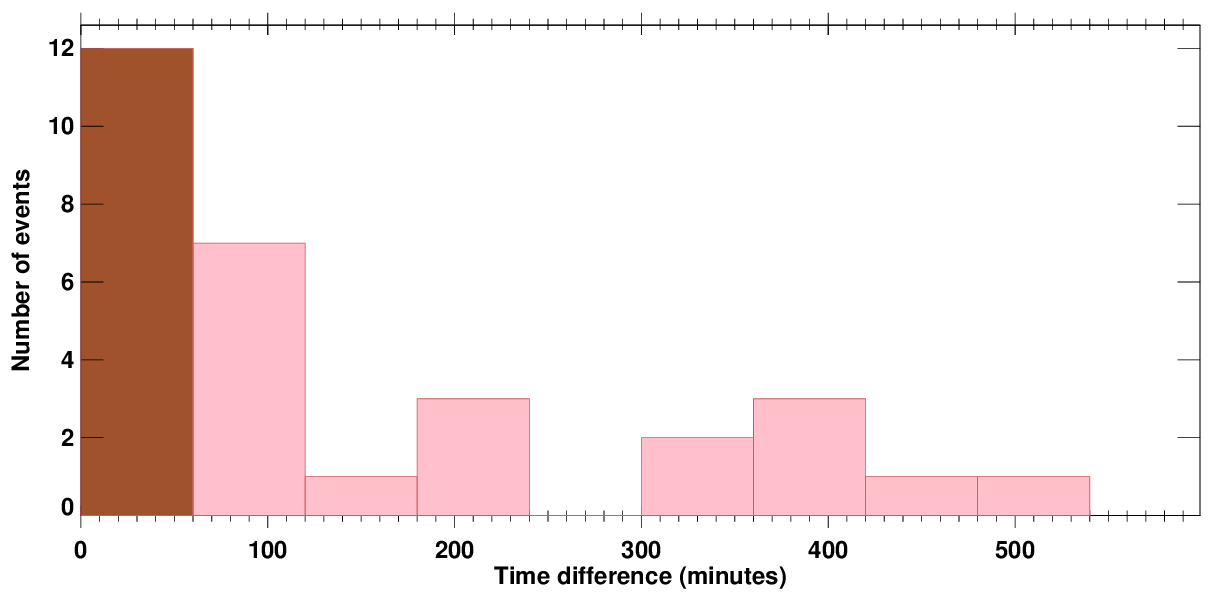}
\caption{Histogram of the number of events versus the time difference between
the onset of the eruptive flare and the appearance of the HFR.}
\end{figure}

\section{Classification of events}

The first and last milestones in the evolution of an eruptive flux
rope in the low corona are the time of its appearance and the time  of
its eruption, respectively.  The  appearance times of the HFRs were
determined using the morphological criteria outlined in Sect. 2 (see
also Paper I). We note that an uncertainty  of one minute in
the determination of the formation times of the HFRs is introduced by
the cadence of the datasets we used. However, this should be 
considered as a lower limit; in fact, an uncertainty greater 
than the nominal 1 minute should be present due to the ambiguities involved
in determining the appearance of the HFRs in the 131 \AA\ data; this
was done visually and is inevitably subjective.

An accurate determination of the onset time of an eruption would
require the construction of a height-time plot computed along the
central direction of the ejection as well as its subsequent fitting with
some function; this could capture the kinematic phases of the erupting
flux rope (e.g., see Zhang et al. 2012; Cheng et al. 2013, 2014a,
2020) and, particularly, the break point, which indicates the onset time
of the main acceleration phase of the eruption. We did not follow that
approach because of the large size of our dataset and the complexity
of the relevant calculations (see the discussion in Cheng et al.
2020). We also did not use results from the backward extrapolation of
white-light CME height-time plots because they do not reflect the
early  evolution of the event. Instead, we used the onset time of the
eruptive flare as a reasonable  proxy for the onset time of the
eruption. Following Cheng et al. (2020), we found the onset
time of the eruptive flares by inspecting the GOES 1-8 \AA\ flux time 
profiles and determining the time beginning from which the flux increases
monotonically and far more rapidly than before. In 30 out of the 34
events, our estimates agreed with the flare onset times provided by
the GOES flare list\footnote{https://www.ngdc.noaa.gov/stp/space-weather/solar-data/solar-features/solar-flares/x-rays/goes/xrs}, while in the remaining cases the differences
were less than 5 minutes.  Cheng et al. (2020) find differences
between the onset time of the flux rope's acceleration phase and the
onset of the eruptive  flare that ranged from -4 to 2 minutes for the
six active region 131 \AA\ hot channel structures they
studied. 
Therefore, our approach may introduce an uncertainty of about
3-4 minutes in the determination of the onset time of the eruption.

Overall, the HFR was formed during the eruption in six events and
appeared  before the eruption in 28 events (we note that in four out of
the  28, the formation time was ambiguous). However, this simple
grouping does not  provide much insight into the situation. We refined
our classification by considering whether or not the appearance of the HFR
was associated with the occurrence of a confined flare. We
define an enhancement in the 131 or 94
\AA\ light  curves with the following properties as a ``confined flare'' : (i) its maximum
intensity is above the  1.5$\sigma$ level of the background emission,
and (ii) it drops below half its peak before the onset of the eruptive
flare.  

The choice of the $\sigma$ multiplication factor used in our
definition of  confined flares certainly affects the results. Our
experience with the data shows that a threshold at about
2$\sigma$ or higher would miss a few cases with conspicuous brightenings in the
hot AIA passbands, while a threshold lower than 1.5$\sigma$ would
tend to include weak transient activity (e.g., Shimizu 1995; Hannah et
al. 2008), which is ubiquitous in active regions.  With our second
criterion for the definition of confined flares, we excluded  precursor
brightenings that occurred in close temporal proximity with large
eruptive flares. 

According to our refined scheme, we classified our events into the following 
five categories. First are HFR events that are formed on the fly during the eruption with
no evidence of prominence material in the 131 \AA\ 
images (FLY in Table 1).  The HFR appears at or after the onset of the 
eruptive flare. Examples are presented in Figs. 1-3.
The second category is the same as the first group but prominence material can be
identified in 131 \AA\ images (FLYP in Table 1). An example is presented
in Fig. 1. Third are events with HFRs that are formed before the onset of the
eruptive flare but  whose formations are not associated with any
distinct confined flares (PRE in Table 1).  Examples are presented in
Figs. 4-6.
 The fourth category is the same as the third group, but the formation is associated with a confined flare
(PREC in Table 1). Examples are presented in Figs. 7-11.
Fifth is uncertain events in which the flux rope is most probably formed
before the eruption and where there is ambiguity in the determination of
its appearance; this is primarily because more than one appear during the
eight-hour interval and it is unclear which one finally erupts (UNC in 
Table  1).

In ten regions that produced events very close to the limb, we detected
partially occulted confined flares during parts of the eight-hour-long
intervals that we used. However, post-flare loops
were  readily visible beyond the limb  in all cases. In the hot 94 and 131
\AA\ passbands, the emission from post-flare loop arcades is usually
higher than the footpoint emission; therefore, it is rather unlikely
that we missed, due to partial occultation, the registration of any
significant confined  flare activity in the 94 or 131 \AA\ light
curves. 
Furthermore, since flux ropes correspond to presumably longer
magnetic field lines that expand above post-flare loops, we are
confident that we did not miss any early formation  of flux ropes
associated with occulted confined flares.

Our classification is presented in Table 1. The first three columns
(from left to right) show the event number as was registered in Paper
I, the start and peak times of the eruptive flare, and the time of
appearance of the HFR. In the fourth column  we give the
time difference, $\Delta t$, between the onset of the eruptive flare
and the appearance of the HFR. In the fifth column we provide
information about the  possible occurrence of a confined flare
associated with the appearance of the HFR.  In parentheses we give the
maximum value of the 131 \AA\ confined flare emission in million data
numbers per second  (DN s$^{-1}$) and, when there was also trace of
the flare in GOES 1-8 \AA\ flux time profiles, its strength according
to its GOES classification. We note that we consider the 131 \AA\ values more
reliable due to the higher sensitivity of the AIA and because the
GOES light curves record full-disk fluxes. Events for which the
confined flare was detected only in the AIA data are marked with an
asterisk.  In the sixth column we give the classification of the
events into five categories.


\section{Formation patterns of eruptive hot flux ropes}

\subsection{Hot flux ropes formed during eruption}

In the left-hand column of Fig. 1, we present an example of a FLY
event, while in the middle and right-hand columns we present examples of
FLYP events. In the  bottom row of the figure, we show the three
HFRs that are formed during the  eruption. The flare brightenings are
prominent in all three bottom panels, while the HFRs are marked with
arrows. In panel (d), the HFR appears as a rounded arc coexisting with
some prominence material; there is also a thin elongated
current-sheet-like structure underneath the flux rope. In panel (b), the HFR appears as an elliptical blob that sits on a $\Lambda$-shaped
loop-like structure, while in panel (f) its presence is  reflected by
the twisted 
shape of the loops. In the same figure, the
corresponding top panels show characteristic snapshots of the
preeruptive  configuration a few minutes before the onset of the
eruption. No conspicuous  trace of HFRs or of brightenings associated
with their appearances  can be detected in the pre-event images, and
that was the case throughout the eight-hour-long intervals that we
checked. However, prominence material can be seen in both panels (c)
and (e), which could indicate the presence of preexisting flux ropes
(if we assume that prominence plasma can be found in flux rope
segments), which are not hot due to insufficient heating through
magnetic reconnection. 

A more detailed presentation of a FLY event (event 89 of Table 1) is
given in  Fig. 2 (see also the associated movie). Panels (a)-(g) show
selected snapshots of the evolution of the pre-event configuration in
131 \AA. The active region  consists of semi-circular loops, which are
not static; no major restructuring takes place for more than 9.5
hours, although weak transient brightenings can occasionally be
detected. Starting at 17:24 UT (panel (h)), bright sheared loops appear
intermingled with lower-lying prominence-like material. These loops
rise, grow, and tangle along their spine as shown in panel (i), which
probably marks the time of appearance of the HFR. Around the same
time,  the core of the active region flares up. Panels (j) and (k)
show the full development  of the rising HFR, while in panel (l) the
HFR is marginally detectable due to its proximity to the edge of the
field of view. 

Panels (j)-(l) show that the HFR ejection occurs simultaneously with
the  eruptive flare. This is also shown in the light curves of Fig. 3,
which have been computed from the region enclosed by the box of
Fig. 2, panel (j). In Fig. 3, the vertical dashed line marks
the appearance of the HFR. We note the precursor local peak in the
94 \AA\ light curve about 30 minutes before the appearance  of the
HFR. That local peak is related to a diffuse transient low-lying
enhancement, which is marginally visible at 131 \AA,\ just above the
limb in Fig. 2, panel (f). In any case, according to our criteria (see Sect.
3), it does not qualify as a confined flare because its emission has not dropped below its half
maximum value at the onset
time of the eruptive flare.

\subsection{Preexisting hot flux ropes without confined flares}

Snapshots of two events with preexisting HFRs whose formations were not
associated with the occurrence of confined flares (PRE events) are
presented  in Fig. 4. For both events, the top row corresponds to the
early development of the HFRs, while the images of the bottom row were
obtained at a later stage. In both events, the formation of the HFRs is
not associated with any confined flaring activity, as was defined in
Sect. 3.  In the event presented in the left-hand column, the HFR appears
as a twisted 
structure relatively high above the limb
(uncorrected height of about  180\arcsec, see  panel  (a)). Sixteen minutes
later, in panel (b), the HFR has moved outward and its size has
increased, while lower down bright flaring loops have appeared. In the
event presented in the  right-hand column, the HFR appears as a round blob
four minutes before the onset of the eruptive flare (see panel (c) for a
characteristic snapshot of its early development). Panel (d) shows
that six minutes later the HFR has grown, moving in the southeast
direction while the eruptive flare is in progress. 

In Fig. 5 (see also the associated movie), we give a more detailed
presentation  of a PRE event (event 61 of Table 1). Panels (a)-(d)
show selected snapshots from the preeruptive configuration, which show
no evidence of the presence of a flux-rope-like structure (see also
the movie). From about 00:34 onward (see panel (e) for an example), the
structure of the active region changes because of the development of
undulatory loops, which grow and gradually develop tangled threads and
twisted tips that probably both reflect the appearance of the HFR
(panel (f)). Some of  the twisted tips merge to form a round blob  that
sits on top of a $\Lambda$-shaped system of loops (see panels (f) to (i)).
The development and outward motion of that feature can be tracked in
panels (j)-(k), while at the same interval the eruptive flare is in
progress. In panel (l), the HFR has left the field of view of the
instrument. 

The light curves of Fig. 6 have been computed from the region inside
the box of Fig. 5, panel (j). Both the 131 \AA\ and 94 \AA\ (and to some
extent the 335 \AA) light curves show enhancements that start around
the time of appearance of the HFR and are practically
indistinguishable from the main eruptive flare. 

\subsection{Preexisting hot flux ropes with confined flares}

In the events with preexisting HFRs that were formed in conjunction
with confined flares (PREC events), the flare appeared in most AIA
passbands that we studied, and occasionally in the X-ray fluxes
recorded by GOES. The confined nature of the flares was judged by the
absence of large-scale EUV dimmings in the AIA data and the absence of white-light CMEs
in the Large Angle Spectroscopic Coronagraph (LASCO; Brueckner et
al. 1995) data.

In Fig. 7, we present snapshots of three PREC events. The top row shows
the HFRs a few minutes after their appearance, while the bottom row
shows them in the course of the eruption. The HFRs of the left-hand,
middle, and right-hand columns (events 81, 97, and 133 of Table 1)
appeared 197, 86, and 451 minutes, respectively, before the onset of
the  eruptive flare. At the early stage of their development, the
HFRs of events 81 and 97 (see panels (a) and (c)) show ring-like structures 
(we note that the flux-rope-like feature south of the  HFR marked with the arrow 
in panel (a) did not participate in the eruption) that sit on top of thin 
elongated current-sheet-like structures, and this shows  better for event 
81. On the other hand, panel (e) indicates that the HFR of event 133
appears as a round blob that sits on top of a $\Lambda$-shaped loop.
Although the time interval between the images of panels (a) and (b) is 190 
minutes, the HFR morphology and size has not changed significantly.
However, panels (d) and (f) show that the HFRs of events 97 and 133 
have grown considerably in the course of their evolution; in addition to its
growth,  the south part of the initial ring-like structure
has deformed in event 97.

In Fig. 8, we present the evolution of a PREC event (event 25 of Table
1)  consisting of an HFR seen face-on (see also the associated movie).
The pre-event configuration appears in panels (a) and (b). From about
08:20 UT, preflare loops rise  and the core of the active region brightens
(panel (c)). These loops grow and develop tangled threads that mark the
appearance of the HFR (panel (d)) while the confined flare progresses
(panels (e)-(g)). As the confined flare decays (see also the light curves
of Fig. 9), the HFR becomes fainter because it presumably cools. It can be 
traced until about 09:40 and then reappears shortly after 10:15 (panel (j)), 
a few minutes before the onset of the eruptive flare (panels (k) and (l)). The
eruptive flare occurred 98 minutes after the appearance of the HFR. This behavior 
is very similar to the evolution of the HFR in Patrsourakos et al (2013).

Event 25 was also presented in Paper I, where we had missed the early
development of the HFR (we noted that ``The first evidence of the hot
flux rope appears around 10:24.'') because we used a 55-minute-long
sequence of 131 \AA\ images  that covered the interval from 10:09 to
11:04. This demonstrates that the use of rather limited time intervals
in the study of flux rope formation times may lead to inaccurate
conclusions.

A PREC event that appeared even earlier (356 minutes), with respect to
the eruptive flare, than event 25 is presented in Fig. 10 (event 118
of Table 1; see also the associated movie). Panel (a) shows
the pre-event configuration. The first detection of the flux rope was
made after 21:43 in conjunction with the development of a
confined flare (see panel (c)). Although the GOES light curve does not
show the flare onset (probably because of  the elevated background level),
it appears clearly in all AIA channels that we studied  (see the light
curves of Fig. 11). The early development of the flux rope (panels
(c)-(f)), which consists of a blob on top of the flaring loops (i.e., its
morphology was consistent with a flux rope seen edge-on; see Sect.
2), follows the development of the confined flare, which shows a
prominent peak 40 minutes after its onset. During the decay phase of
the flare, the flux rope cools down and becomes fainter until it cannot
be traced in AIA  images (panels (g)-(j)). In the extended interval
between the end of the confined flare and the onset of the eruptive
flare, the flux rope reappears for short periods, centered around 23:41,
01:31, 01:53, and 03:39 (see the associated movie), which coincide with
local  peaks in the light curves of the 131 \AA\ and 94 \AA\ emission
(see Fig. 11).  We conjecture that flaring activity provided plasma
heating through magnetic reconnection. Unfortunately, due to the
location of the HFR close to the limb, we cannot check whether
reconnection also  changes the magnetic flux of the flux rope, as has
been found by Guo et al. (2013). The flux rope appears again just
before the eruptive flare, in the course of which it rises (see panels
(k) and (l)) until it leaves the field  of view of the instrument. 

In the description of events 25 and 118, we mentioned the existence of time
intervals in which the flux ropes ``disappeared'' from the 131 \AA\ images.
The same was also true for their 94 \AA\ emissions. Such time intervals were
found in all PREC events at both 131 and 94 \AA. These intervals start from
about 20 minutes to more than 1.5 hours after the appearance of the PREC 
flux ropes and can last from about 30 minutes to more than 3.5 hours. Cooling 
of the flux ropes is a possible explanation for such behavior. The varying
onset times of these intervals with respect to the appearance of the flux
ropes may reflect different occurrence times of reconnection events, which
could power the confined flare and heat the plasma. 

Cargill (1994) assumed that  loops impulsively heated to flare temperatures 
cool initially by conduction and at later times by radiation. 
The time scale 
of cooling due to thermal conduction is

\begin{equation}
\tau_c = \frac{21 n k L^2}{18.4 \times 10^{-7} T^{2.5}}
,\end{equation}
where $k$ is the Boltzmann constant and $n$, $T$, and $L$  are the number
density,  temperature, and characteristic length scale of the
structure, respectively (e.g., Pagano et al. 2007). On the other hand,
the time scale of radiative losses is 

\begin{equation}
\tau_r = \frac{3 k T}{2 n P(T)}
,\end{equation}
where $P(T)$ is the radiative loss function. Results for flux rope
plasmas  with different values of $n$ and $T$, under the assumption of
$L  \approx 30$  Mm (e.g., see Cheng et al. 2012; Patsourakos et
al. 2013), are presented in  Table 2. The value of $n=12 \times 10^9$
cm$^{-3}$ has been taken from Syntelis  et al. (2016),  who performed
differential emission measure 
analysis of two HFR  events observed by
the Hinode EUV Imaging Spectrometer (EIS)
and AIA, while the value of $n=1.1 \times 10^9$  cm$^{-3}$
is the average number density of HFRs derived by Cheng et al. (2012)
using AIA data. A temperature of 8 MK is appropriate for HFRs visible
in the 131 and 94 \AA\ passbands for which the radiative loss function
can be approximated by $P(T)=5.49 \times 10^{-16} T^{-1}$ (Klimchuk et
al. 2008). The value of $\tau_c=16$ min  deduced from the large number
density appears  broadly consistent with our observations, while both
values of $\tau_r$  (i.e., 33 and 348 minutes) should be considered as
upper limits because they  do not take into account the  conductive
cooling of the plasma, which presumably occurred earlier. To this end,
we also provide the values of $\tau_r$ for a lower
temperature of 2.5 MK in Table 2. For that temperature, the approximate expression
for $P(T)$ becomes $P(T)=3.53 \times 10^{-13} T^{-3/2}$ (Klimchuk et
al. 2008), and we obtain values of $\tau_r$ that are a factor of about
4 lower than before.

\begin{table}[h]
\begin{center}
\caption{Cooling time scales of hot flux rope plasmas}
\label{ALMAobs}
\begin{tabular}{lccc}
\hline 
Density & Temperature & $\tau_c$\tablefootmark{a} & $\tau_r$\tablefootmark{b}  \\
(10$^9$ cm$^{-3}$) & (10$^6$ K)  & (minutes)           &  (minutes) \\
\hline
12.0              & 8           & 16                  & 33 \\
1.1               & 8           & 2                  & 348  \\
12.0              & 2.5         & -                   & 8  \\
1.1               & 2.5         & -                   & 84 \\
\hline 
\end{tabular}
\tablefoot{The characteristic flux rope length scale is 30 Mm. 
\tablefoottext{a}{$\tau_c$ denotes the time scale of cooling due to thermal
conduction.}
\tablefoottext{b}{$\tau_r$ denotes the time scale of radiative losses.}
}
\end{center}
\end{table}

The cooling at the site of the confined flare is evident in the light
curves of some PREC events (see Fig. 11 for an example), where the
flare peak generally tends to appear progressively in passbands that
probe cooler material (see Viall \& Klimchuk 2012, 2013). In five of
these  events, the flux rope also appears, for limited time intervals,
after the confined flare in some of the  cooler  passbands centered at
211 and 335  \AA. Those were the PREC events whose formations were
associated with the  strongest confined flares, as judged by the
maximum value of their intensity at 131 \AA\ (see the relevant entries
in the fifth column of Table 1).  As a heating event becomes stronger,
more plasma is evaporated and  therefore one could anticipate
higher densities of the cooling plasma.  On the other hand, there was
not a single case in which we  could track the  flux rope in cooler
passbands throughout the interval of its disappearance  from the 131
and 94 \AA\ images. This could be attributed to the low density
of the flux rope material (see Syntelis et al. 2016). The difficulty
in observing cooling flux rope plasmas at temperatures of a few
million K may also be related to the fact that during periods without
flare activity, the active region differential emission measure 
peaks
in that temperature range, thus making it possibly more difficult to
detect the cooling of the flux rope plasma against a higher background.

\begin{table}[h]
\begin{center}
\caption{Statistics of the patterns of HFR formation}
\label{ALMAobs}
\begin{tabular}{lcc}
\hline 
Classification\tablefootmark{a} & Range of $\Delta t$\tablefootmark{b} & Number of events  \\
                         & (minutes)           & \\
\hline
FLY               &  [-5, -2]              & 3 \\
FLYP     & [-3, 0]    & 3 \\
PRE               & [3, 39]              & 4 \\
PREC              & [51, 536]              & 20 \\
UNC               & -           & 4 \\
\hline 
\end{tabular}
\tablefoot{
\tablefoottext{a}{The acronyms FLY, FLYP, PRE, PREC, and UNC are 
explained in Sect. 3.}
\tablefoottext{b}{$\Delta t$ denotes the time difference between the onset of
the eruptive flare and the appearance of the HFR.}
}
\end{center}
\end{table}

\section{Discussion and conclusions}

This paper presents the first statistical survey of EUV observations
to  determine the formation times of eruptive HFRs relative
to the CME  initiation. We considered 34 M-class and X-class flares
that occurred relatively close to the limb and involved hot
channel or hot blob  configurations, which were interpreted as
HFRs in Paper I. Using uninterrupted sequences of 131 \AA\ images that
spanned more than eight hours, we determined the formation time of the
HFRs using the same morphological criteria that we had used in Paper
I. Our results are summarized in Table 3 and in the histogram of
Fig. 12. Our main  conclusions are as follows.

(1) Two thirds (20/30) of the events for which the formation time of
the HFRs  was unambiguously determined involved a preexisting HFR
(that is, an HFR  that was formed before the onset of the eruptive
flare) whose formation was  associated with the occurrence of a
distinct confined flare (PREC events).  There were four events with a
preexisting HFR that was not formed in conjunction  with a confined
flare (PRE events) and six events in which the HFR was formed  once the
eruption was underway In the latter group, there was
no evidence of prominence material in the 131 \AA\ images
(FLY events) in three of the events, while in the other three events there was (FLYP events).

(2) In the FLY and FLYP events, the time difference, $\Delta
t$, between the onset of the eruptive flare and the appearance of the
HFR is either zero or negative. After setting these negative
values to zero, we found that the mean and median values of $\Delta t$ for the
whole population of events were 151 and 98 minutes,  respectively.

(3) The time difference, $\Delta t$, was larger, in an average
sense, in   events that involved a PREC flux rope than in those
that involved a PRE  flux rope; in the former group, $\Delta t$ ranged
from 51 to 536 minutes, while  in the latter it ranged from 3 to 39
minutes. Our sample is too small to know if we have two populations in
a statistical sense. However, there is evidence that the formation
patterns of HFRs either during or before eruptions do not show
substantial  differences when the appearance of the flux rope is not
associated with any  distinct confined flare activity. This conclusion
is further reinforced if we take into account the uncertainties of 3-4
minutes and at least one minute (the latter being a lower limit)  
that are associated with the determination of the onset of the eruptions 
and the  formation times of the HFRs, respectively (see Sect. 3). The 
segregation of the PREC events is also reflected in the histogram of Fig. 12; 
all but two of the 12 events that populate the leftmost bar of the histogram 
(brown color) are either FLY, FLYP, or PRE events, while the 
events that populate all the other bars (pink color) are exclusively 
PREC events.

(4) The existence of a significant population of preexisting HFRs that were
formed well before  the initiation of the eruptions ($\gtrsim$1 hour) is 
consistent with previous case studies on the formation times of sigmoidal
EUV hot channel structures that were located close to disk center. These 
studies include Cheng et al. (2015), Chintzoglou et al. (2015), Joshi et al.
(2015), Zhou et al. (2017), James et al. (2018), and Wang et al. (2019), who
reported hot channel formation times relative to the onset of the eruptions of 
about 3-4.5, 11, 1.5, 2.5, 2, and 5 hours, respectively. 

(5) We emphasize that the use of uninterrupted, long (more than 8 hours) 
sequences of images was essential for revealing the statistical prevalence
of ``truly'' preexisting HFRs that were formed well before the onset of
the eruptions. For example, if we had used one-hour-long sequences of images,
as in Paper I, we would have missed the formation time of 18 HFRs because they
appeared more than one hour before the onset of the corresponding 
eruptions. We also speculate that revisiting published case studies of HFRs 
that employed rather short sequences of images may also shift the formation of 
some flux ropes to earlier times with respect to the initiation of the 
eruptions (see also Sect. 4.3). 

(6) It is possible that the percentage of truly preexisting HFRs
that we  reported are lower limits of the actual ones for the
following reasons. (i) In three out of the six events with HFRs formed
on the fly, there is evidence of prominence  material
in 131 \AA\ images, which may indicate the presence of preexisting
flux ropes (FLYP events of Sects. 3 and 4). Therefore, the only
difference between such events and the events with preexisting HFRs
may be the amount of heating available prior to the eruption. (ii)
In all four events grouped as ``uncertain'' in Sect. 3, the HFRs were
formed well before the onset of the eruption, but we were unable to
unambiguously determine the time of their appearance. (iii) Only three
PREC events contained a flux  rope configuration seen
face-on. Although this percentage (3/20) is similar to  the percentage
of flux ropes seen face-on that was reported in Paper I (9/45), we
cannot exclude the possibility that, due to their poor visibility
compared to flux ropes seen edge-on (see Paper I), we missed the early
formation of some HFRs  that initially appeared face-on.
      
(7) Not all confined flares recorded in the light curves we constructed 
were associated with the formation of flux ropes. It is difficult to establish 
a pattern or common characteristics for those that do, but this is a
topic that deserves further investigation.

(8) Our results provide, on average, 
indirect support for CME initiation models that involve preexisting magnetic flux
ropes that subsequently erupt, although for a minority of cases models
in which the  flux rope is formed during the eruption might be more
appropriate.  Due to the lack of suitable magnetic field data,
it is difficult, and beyond the scope of this paper, to  address 
the question about which events are initiated by ideal mechanisms and then
pinpoint the relevant instability, or to speculate about the possible 
contribution of the tether-cutting scenario versus the breakout model. For 
limb events such as the ones we  studied, these questions could be
addressed in the future by combining the AIA  observations with
simultaneous vector magnetograms of the source region from the recently 
launched Solar Orbiter mission in quadrature or from an L4 or L5 mission.

\begin{acknowledgements}
We thank the referee for his/her constructive comments.
A.N., X.C., and J.Z. thank the ISSI team on ``Decoding the Pre-Eruptive
Magnetic Configuration of Coronal Mass Ejections'', led by S. Patsourakos and 
A. Vourlidas, for stimulating our research. AV is supported by NSF grants 
80NSSC18K0622 and NNX17AC47G. X.C. is supported by NSFC grants 
11722325, 11733003, 11790303, 11790300, Jiangsu NSF grant BK20170011, and the 
Alexander von Humboldt foundation. J. Z. is supported by NASA grant 
NNH17ZDA001N-HSWO2R.

\end{acknowledgements}

\end{document}